# Blockchain-Based Applications in Higher Education: A Systematic Mapping Study

Bakri Awaji
School of Computing
Newcastle University
Newcastle, UK
b.h.m.awaji2@ncl.ac.uk

Ellis Solaiman
School of Computing
Newcastle University
Newcastle, UK
ellis.solaiman@ncl.ac.uk

Adel Albshri
School of Computing
Newcastle University
Newcastle, UK
a.albshri2@ncl.ac.uk

## ABSTRACT
The utilisation of blockchain has moved beyond digital currency to other fields such as health, the Internet of Things, and education. In this paper, we present a systematic mapping study to collect and analyse relevant research on blockchain technology related to the higher education field. The paper concentrates on two main themes. First, it examines state of the art in blockchain-based applications that have been developed for educational purposes. Second, it summarises the challenges and research gaps that need to be addressed in future studies.

## 1. INTRODUCTION
Within the higher education sector, the traditional system of education has seen significant and continuous improvement through the application of new technology such as the Internet, and the World Wide Web. Web-based applications are being applied to improve communication, increase collaboration, sharing resources, and promote active learning. Blockchain applications for education are relatively new, and the number of products based on blockchain is currently small. However, Blockchain has the potential to provide many new opportunities.

Blockchain was first developed for the Bitcoin digital payment system in 2008 [2]. This emerging technology has since grown rapidly and has become the subject of intense research in many industries, research organisations, and universities around the world [3][8][49]. Blockchain aims to solve the problem of a "trusted" central authority having responsibility for mediating transactions between different parties. Centralisation can lead to security problems such as being a single point of failure, as well as other problems such as cost. The decentralised nature of blockchain improves trust between parties in a system and eliminates the need for a trusted third party to perform transactions [7]. As a distributed database, blockchain stores every transaction between the parties. Within a blockchain network, a ledger that maintains a record of all transactions is replicated and shared with all parties.

Blockchain is also capable of running smart contracts (executable code) [7]. Smart contracts increase the effectiveness of blockchain solutions and allow for distributed applications to be deployed in numerous fields for various purposes. Within the education sector, smart contracts can be used to build flexible blockchain based distributed solutions for the benefit of all participants in an online learning system, including students, teaching staff, and administrative personnel. For example, it may become possible for students and education institutions to contract more personalized digital agreements which specify assignment requirements, time frames, and grading structures [8].

There are a number of blockchain technology features that make it worthy of investigation for enhancing educational systems:

- Immutability: the data stored on the blockchain is tamper-proof due to the chronological order that data is stored and the cryptography that secures and connects blocks.
- Reliability: The decentralized nature of the network means that it operates in a more reliable fashion than centralized systems. There is no central authority that could fail.
- Transparency of information is a growing demand. With blockchain technology, it is possible to create highly transparent decentralized data storage.
- Availability: the distributed nature of blockchain infrastructure means that data is replicated, stored closer, and accessed more efficiently by owners of the data.
- Trust: Blockchain technology eliminates the need for a trusted third-party service provider to enable communication between parties.

Therefore, the main aim of this study is to investigate the research topics that have been examined in the field of education within the context of blockchain and smart contract technologies, and to identify important challenges that need to be addressed in future studies. To achieve this aim, we selected a systematic mapping study as our methodology. We followed the systematic mapping process presented in [7]. We searched for relevant papers in the scientific databases and then produced a map of the current blockchain-related applications for education. This map contains important information that may help to understand the issues and challenges concerning blockchain for future research projects.

The remainder of this paper is organised as follows: Section 2 presents background information on blockchain technology and its features as well as smart contracts and their structure. This section also provides some details on current blockchain projects in the higher education field. The methodology that was adopted in this study is described in Section 3. Section 4 presents the result of the search for relevant papers and the classification of topics related to blockchain in higher education. Section 5 discusses the results of the study and answers the research questions. Section 6 concludes the paper.

## 2. BACKGROUND

This section provides information about blockchain technology and its features. Further, this section highlights examples of current blockchain projects in education. Additionally, it presents a brief explanation of smart contracts and their structure.

### 2.1 Blockchain Technology

Blockchain is a distributed database that stores the transactions sent between the participants in a secure and immutable way. Blockchain is a P2P network that allows nodes (peers) to collaboratively maintain the network for block and transaction exchange [16].

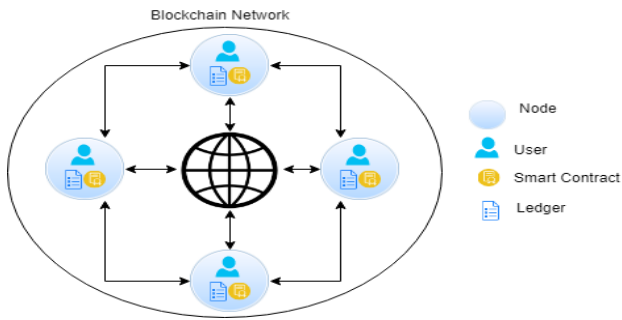

**Figure 1: Blockchain network overview.**

Therefore, there is no need for a third trusted party since the participants can communicate and send transactions between each other directly. A cryptographic hash identifies the block, and each block references the previous one, which creates a chain of blocks (Figure 1). Each block contains several transactions, and the maximum size of each block varies according to the blockchain platform type, for example, 1 Mb in Bitcoin and between 20 to 30 kb in Ethereum [42]. The blocks in the chain are immutable and cannot be changed, which prevents the double-spending problem [50].

Historically, the first generation of blockchain was cryptocurrency, which is a digital currency based on cryptography and P2P networks. One of the concepts integrated into blockchain is mining, which is the process of adding transaction records to the public ledger of past transactions on the blockchain. The mining process requires a miner on the network to generate the new block of transactions by collecting those transactions into a block, running a mathematical process to verify the block and adding it to the chain of past blocks.

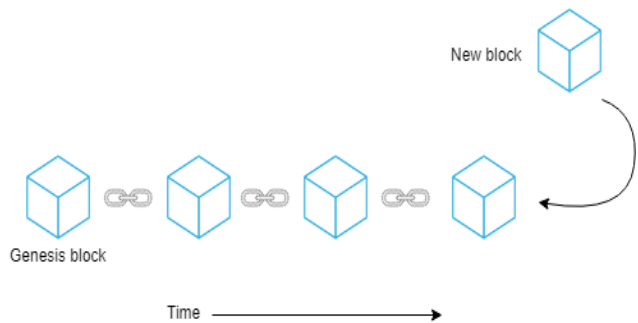

**Figure 2: Adding a block to the chain.**

The other miners' nodes in the network can validate the newly generated blocks through a consensus algorithm. The second generation of blockchain has emerged in the form of Ethereum [8], which allows for building and implementing distributed applications. The Ethereum blockchain allows smart contracts to be built on top of it, and this has opened the door for researchers to integrate blockchain into various fields.

Blockchain has generally been divided into two main types: public and private blockchain [8]. Public blockchain (such as Ethereum) allows anyone to join and participate in the network. In contrast, in a private blockchain (such as Ripple), only users with permissions can join and participate in the network.

### 2.2 Smart Contracts

A smart contract refers to an event–condition–action stateful computer program that is carried out between two or more parties who do not have implicit trust with one another [28]. In other words, it is a self-executed code that is run to apply roles and conditions between two or more parties [2]. By applying a smart contract using blockchain technology, there is not only a reduction in third-party costs within the transaction process, but there is also improved transaction security. A smart contract can be either centralised or decentralised; it can be implemented to run off-chain in a centralised environment or to run on blockchain in a decentralised environment [29][51]. Figure 3 presents the two types of smart contracts.

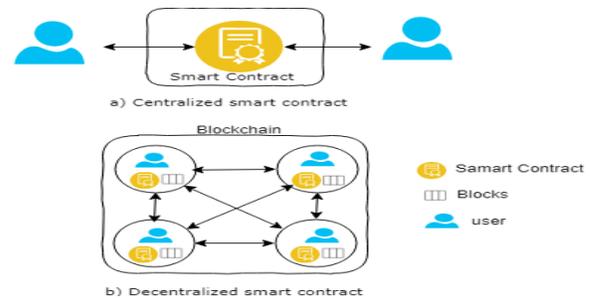

**Figure 3: Smart contract types.**

### 2.3 Blockchain and Smart Contract Applications within Education

In recent years, the role of blockchain applications in education has received increasing attention across several disciplines. Blockchain and smart contract technologies are increasingly involved in education in different means and forms. However, until now, blockchains in education have been primarily used to record grades and award certificates; far too little attention has been given to the utilisation of blockchains and smart contracts for building an infrastructure for the learning process [4]. Some of the current blockchain and smart contract applications within education are as follows:

- Digital certificate applications: These applications are intended to provide greater control over the students' earned certificates and to decrease dependence on third-party intermediaries – including employers and universities – for storing, verifying and validating credentials. Examples are Open Blockchain [8] and the Blockcerts project [9].

- Support services applications: These applications are aimed at establishing a particular cryptocurrency based on Bitcoin for regulating the market of goods and services in the educational field, such as support services, regulated studies, enrolling in online courses and training institutions' micro-contracts with the digital transaction of economic assets to acquisition books. An example is Edgecoin [33].

- Earnings applications: These Applications connect learning with earnings. In this case, the blockchain is used for storing the teaching or learning hours and not the digital currency. An example is the Ledger project [44].

There are other potential applications for blockchain and smart contracts within education, such as for student assessments, online learning, finance and payment, digital rights management, distributed file storage and identity management.

## 3. RESEARCH METHODOLOGY

In this study, we used the systematic mapping method described in [34] to explore blockchain applications related to education (Figure 4). The results of this study may help researchers to identify the gaps and challenges that need to be addressed in future studies.

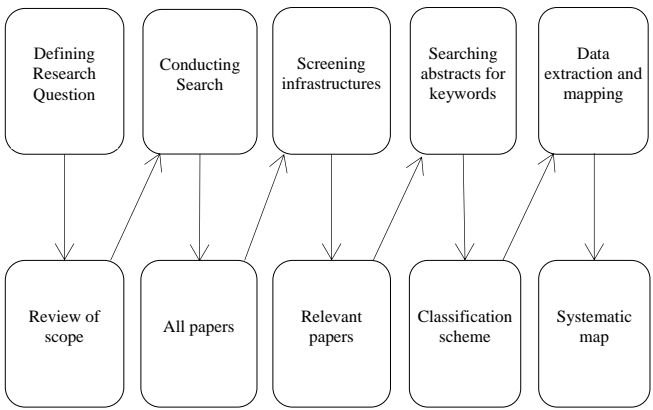

**Figure 4: Steps of the systematic mapping study.**

### 3.1 Definition of Research Questions

In this step, we identified the research questions that our study aims to answer. For our study, we defined the following research questions:

RQ1: What are the research topics on blockchains for higher education?

RQ2: What are the research challenges for blockchains in relation to higher education?

### 3.2 Conducting Search

In this step, we conducted a search for papers relevant to the research topic: blockchain applications in education. To narrow the focus of our study, we selected the terms 'blockchain' and 'education' as the keywords. Furthermore, we used well-known scientific databases to conduct our search: IEEE Xplore, ACM Digital Library, ScienceDirect, Springer and Scopus. These selected scientific databases index high impact, high quality papers in the fields of education and information technology. Table 1 presents the query strings used to search each database. We focused only on high-quality papers that had been published in conference proceedings, journals, workshops, symposiums and books.

### 3.3 Screening for Relevant Papers

To screen the search results for papers relevant to the study questions, we followed the search criteria identified in [12]. First, we excluded irrelevant papers based on their titles. If the title was unclear, we reviewed the abstract of the paper. Furthermore, we excluded non-English papers, those without full text available, duplicates, newsletters and grey literature.

**Table 1. Search queries.**

| Scientific Database | Query Strings |
|---|---|
| ACM digital library | [All: blockchain] AND [All: higher] AND [All: education] AND [All: university] AND [Publication Date: (01/01/2017 TO 01/31/2020)] |
| IEEE Xplore | (((("Document Title":blockchain) AND "Document Title":higher) AND "Document Title":education) OR "Document Title":university) |
| Springer | 'blockchain AND higher AND education AND OR AND university' |
| Scopus | TITLE-ABS-KEY (blockchain) AND TITLE-ABS-KEY (higher education) OR TITLE-ABS-KEY (university) AND (LIMIT-TO (PUBYEAR, 2020) OR LIMIT-TO (PUBYEAR, 2019) OR LIMIT-TO (PUBYEAR, 2018) OR LIMIT-TO (PUBYEAR, 2017))) |

### 3.4 Searching Abstracts for Keywords

In this step, we classified all the relevant papers by using the keywording technique explained in [12]. We identified the main keywords and contribution of each paper from its abstract.

### 3.5 Data Extraction and Mapping

This process was conducted to collect the required information to address the research questions of this study. Therefore, we designed the review criteria to contain nine elements for reviewing the papers, as shown in Table 2. The review criteria were piloted on a sample of five papers and then applied on the rest of the papers to extract data.

**Table 2. Criteria for reviewing papers.**

| Criteria | Description |
|---|---|
| Title | Title of the paper |
| Author(s) | The author name(s) |
| Paper type | Conference, workshop, journal, or book chapter |
| Paper topic | The topic area and subject |
| Publication date | Publication year |
| Publication location | Country of conference or organisation |
| Paper purpose | Aim of the paper |
| Application implementation | System structure and implementation |
| Challenges | Actual and potential challenges |

## 4. STUDY RESULTS

We obtained a total of 108 articles. In the first stage of the screening process, we removed 47 irrelevant articles based on our criteria. Articles were omitted for two reasons. Because we were concentrating on research of blockchains from a technical standpoint in higher education, we omitted papers that were not about higher education specifically. We also omitted papers that discussed the general aspects of a blockchain. After that, 19 additional papers were discarded as duplicates, resulting in 42 papers. Thus, we analysed 42 papers for our systematic mapping research.

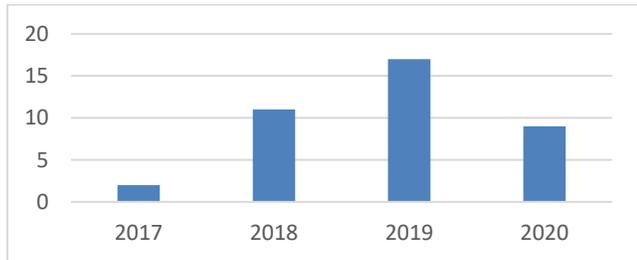

**Figure 5: Number of papers per year.**

Figure 5 shows the distribution of papers by the year of publication. It is important to note that all of the papers were published after 2016. This shows that this is a modern and novel area of research. Thus, the number of publications on this topic appears to be increasing each year, reflecting an increase in interest in the blockchain adoption of higher education.

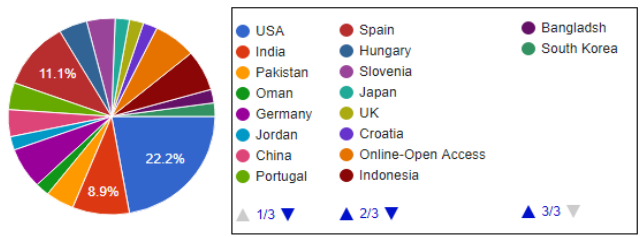

**Figure 6: Geographical distribution of primary papers.**

The geographical distribution of primary papers is shown in Figure 6. This geographical distribution, which is spread across 18 countries, shows that the adoption of blockchains in higher education has received international research attention. The greatest number of articles (n = 10, 22.2%) were written by colleges or corporations in the United States. Spain took second place with three papers (11.1%). Four papers were published in India. The remaining papers were published in other countries.

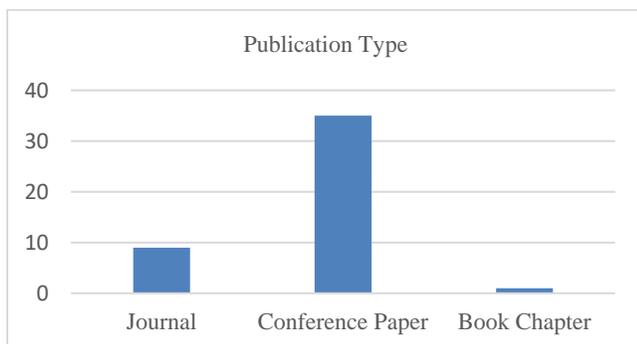

**Figure 7: Distribution of publications by type.**

Figure 7 presents the distribution of publications by type. The vast majority (77.78%) of these articles were published in conference proceedings, and four (20%) were published in journals. The Institute of Electrical and Electronics Engineering (IEEE) had the most publications. Approximately 65.63% of the venues were technical venues, while 34.38% were educational venues.

Table 3 shows the classification categories of the papers. In this step, we classified all the relevant papers by using the keywording technique explained in [12]. We identified the main keywords and classified each paper from its abstract.

**Table 3. Classification categories.**

| Categories | Scientific Papers |
| --- | --- |
| Certificate/degree verification | [4] [6] [10] [14][15][17] [19] [23] [24] [25][36] [41] |
| Student assessments & exams | [9][27][39] |
| Credit transfer | [40][42] |
| Data management | [5] [11][12] [20] [25] [26][31] [32] [35] [38][46][47] [48] |
| Admissions | [13][30] |
| Review papers | [1] [3][5] [8] [11][18] [21] [37][43] [45] |

- **Certificate/degree verification:** This category included studies of blockchain technology that can assist with validating student diplomas and can provide greater control over how students earn certificates.
- **Students assessments & exams:** Articles in this category described automated mechanisms for the production of exams and assessment schemes for university students.
- **Credit transfer:** This category included research on blockchain applications for storing student records and transcripts and transferring academic credits between universities.
- **Data management:** This category contained articles on blockchain applications for connecting students' records across institutions as well as smart contracts for managing student data and storing their records.
- **Admissions:** These articles proposed blockchain applications to facilitate students when applying to universities by storing and sharing the admission procedures and required documents to apply to a particular university.
- **Review papers:** These included literature review studies conducted by researchers during the period defined in this study.

Table 4 presents the distribution of publications by the challenges detected when reviewing the scientific literature included in the search results.

Table 4. Classification of challenges.

| Challenge | Category | | | | |
|---|---|---|---|---|---|
| | Certificate/degree verification | Student assessments & exams | Credit transfer | Data management | Admissions |
| Privacy | [4][15] [41] | [39] | [42] | [5][12] [25][31] | [13][30] |
| Immutability | [43] | [27] | | | [13] |
| Blockchain usability | [15][17][23] | [27] | [40] | [31][48] | |
| Cost | [22][24] | [27] | | [35] | |
| Scalability | [17] | | [42] | [12] [26][35] | |
| Consensus algorithms | [17] [22][24] | [39] | | | |
| Blockchain platforms | [30] | | | [5] | |
| Motivation | [22][24] | | | | [13] |

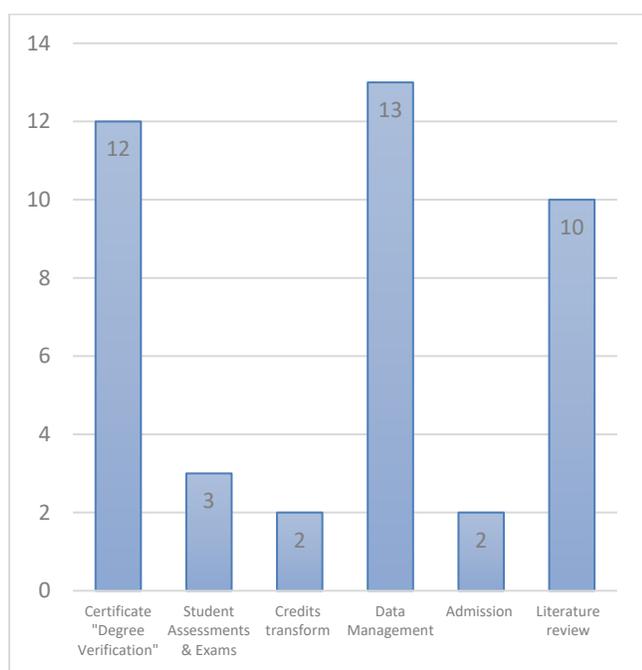

Figure 8: Articles by classification category.

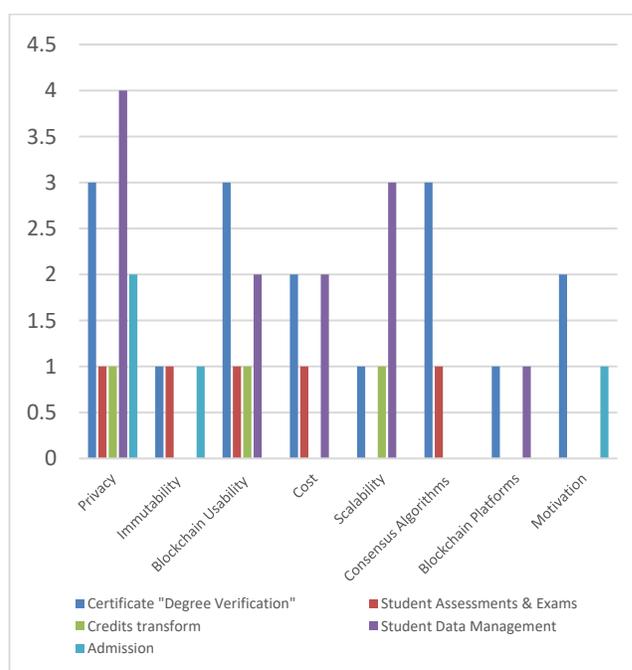

Figure 9: Classification of challenges.

As shown in Figure 8, the majority of studies on blockchain applications in higher education focused on the management of certificates. Of the 42 publications analysed, 12 (28.57%) focused on degree verification. Three papers (7.14%) examined blockchain applications concerning student assessments and exams. The third application type focused on the transfer of credits and included two articles (4.76%). The fourth application type concerned data management and had four articles (30.95%).

Two of the papers (4.76%) focused on blockchain frameworks to ensure security and protection in the admission systems. Finally, the sixth type was review papers that examined potential methods of implementing blockchains in higher education.

Figure 9 illustrates the nine challenges presented in the reviewed papers. The first challenge, with 11 articles, concerned blockchain privacy. These articles examined different privacy concerns that may arise with blockchain technology. Further, immutability (with three articles), a major aspect of blockchains, may become a problem when implementing blockchain technology in higher education. Immutability can make it difficult for educational institutions to enforce new laws on data storage or correct inaccurate data. The third challenge involved the complicated nature of blockchain technology, as stated in seven articles. For example, certification authentication can become a challenge when using blockchain technology in education. The cost of using blockchain technology was the fourth challenge, with four articles. Five articles observed the issue of the scalability of blockchains. As the number of blockchain blocks increases, so does the processing time, which can impact the performance.

Articles four and two focused on consensus algorithms and blockchain platforms, respectively, and noted that educational institutions may face challenges when identifying which data and services need to be introduced in the blockchain network. The final challenge concerned the reasons for using blockchain technology, with three articles noting that immaturity problems, including complicated settings and low usability, have a sustained impact on blockchain's implementation in higher education.

# 5. DISCUSSION

According to the publication trend, the utilisation of blockchain innovation for higher education has been met with increasing enthusiasm. As the studies related to this subject have been relatively few, there is a need for further research examining the potential use of blockchains for education. The evaluation of the 42 studies presented here can help in addressing two major research questions.

## 5.1 First Research Question

*RQ1: What are the research topics on blockchains for higher education?*

Although there have been numerus blockchain-based applications developed for educational purposes, few have been utilised by stakeholders. Such applications can be classified into six categories, as presented in Table 3.

### 5.1.1 Certificate/degree verification
The first category focused on specific applications intended to confirm certificates and verify degrees. This category included all research addressing academic certificates and transcripts for higher education students. Verification of academic certificates is crucial for employers as well as other authorities to confirm the authenticity of an academic degree. Following up with and verifying the status and authenticity of a diploma is difficult, and blockchain technology presents a promising solution [43]. The applications in this category tended to focus on authorising universities to provide students with access to official certificates and ensuring information privacy.

Han et al. [15], for example, introduced a unique blockchain-based method for universities to authenticate and share official education certificates. Budhiraja and Rani [6] also presented the TUDocChain platform, which enables academic certificates to be reliably and sustainably authorised on a public ledger as a practical solution to issue, validate and share certificates.

### 5.1.2 Student assessments & exams
The second category focused on student assessments and exams. These studies intended to fulfil the quality standards as well as requirements established in the relevant programmes and curricula. The researchers also aimed to create blockchains for online quizzes and to fulfil informatisation in higher education. Mitchell et al. [27] introduced dAppER, an automated quality assurance mechanism that uses established internal procedures to develop exam papers and their related evaluation schemes.

This technology offers a robotised quality affirmation system to develop test papers and their appraisal plans, wherein a permissioned blockchain protects the tests and sustains the audits' immutable and trusted ledger. Shen and Xiao [39] proposed a model that uses blockchain technology to verify students' answers.

### 5.1.3 Credit transfer
The third category involved applications for transferring students' credits between universities. These applications are intended to enable a university to transfer completed course credits to another university. In this category, applications were presented to transfer credential records between colleges, establishments or associations, given blockchain's high security and trust. Srivastava et al. [40] proposed a framework to store student records and transcripts that included an electronic credit transfer mechanism. Hence, students are able to transfer completed course credits between different universities. Turkanović et al. [42] also proposed a global higher education credit platform called EduCTX, which follows the European Credit Transfer and Accumulation System.

### 5.1.4 Data management
The fourth category involved data management applications that can gather, report and assess university data to automate rules, support decision making and protect students' identities and their data. Bore et al. [5] examined the initial structure, implementation and assessment of the blockchain-enabled School Information Hub by conducting a case study of Kenya's school system. In addition, Filvà et al. [11] presented a blockchain-based solution for automating rules and constraints so that students can control their data and ensure their privacy and security. Forment et al. [12] also proposed various actions intended to safeguard students' identities and to secure their data through new technologies, including blockchain.

### 5.1.5 Admissions
The fifth category concerned blockchain applications connected to university admissions and student registration. Mori and Miwa [30], for example, presented a digital university admission application system that organises study documents and e-portfolios through smart contracts on a blockchain. Curmi and Inguanez [9] also proposed a platform that registers students and includes their medical records while ensuring that their sensitive data remain private, with the information owner in control of access to these documents. Moreover, Ghaffar and Hussain [13] introduced a system for PEC, HEC and IBCC to verify students' educational records, enabling students to apply for university admissions through a single platform.

### 5.1.6 Review papers
The final category involved literature reviews of studies focusing on higher education and blockchain applications. Overall, eight papers examined this topic. In the literature reviews, preliminary investigations and evaluations were student centred and focused on recordkeeping and sharing using a trusted and safe platform, such as in [1] [8] [18][43] [45].

## 5.2 Second Research Question

*RQ2: What are the research challenges for blockchains in relation to higher education?*

Although blockchains have several beneficial applications for education, researchers continue to face numerous problems in utilising this technology for education. Hence, several challenges were detected in the reviewed papers.

### 5.2.1 Immutability
The term 'immutable' came up repeatedly in relation to blockchains. Blockchain's immutability results in it becoming impossible for the data stored in the blocks to be changed, which is a crucial aspect of blockchain technology. However, immutability is a major challenge to using blockchain technology for education, such as diploma revocation. Although this is not common, it can be used when there are unique circumstances in which diplomas are withdrawn. The diplomas that are stored on the blockchain, however, cannot be changed because of blockchain immutability [43]. Hence, immutability can decrease blockchain's applicability in terms of students' sensitive data. This challenge applies especially to three categories in which students' information needs to be stored on the blockchain: certificate/degree verification, exams/assessments and admissions.

*5.2.2 Blockchain usability*
Blockchain technology's usability is also a major challenge in the education field. The technology's terminology is often unclear, and it is perceived as lacking maturity. Moreover, the user may have to deal with several complicated settings to ensure security, such as primary keys, recovery roots and public keys. It should also be noted that a P2P network blockchain includes very different specifications compared to those intended for individuals, which can make using blockchains difficult for end users. Thus, blockchain usability should be enhanced by application design interfaces, which allow individuals who do not have technical expertise to use blockchain more easily [36]. Thus, further studies on blockchain usability for individuals are needed. In addition, blockchain adaptation can be improved in the education field through well-designed interfaces with simple specifications.

*5.2.3 Privacy*
It is important to consider how data can be securely accessed and used while maintaining privacy [15]. Blockchain systems use private and public keys to protect user identities. Public keys are publicly visible, and thus the blockchain cannot guarantee transactional privacy. Thus, blockchain privacy protection mechanisms have drawbacks that may lead to anonymous abuse. Hence, it is crucial to protect the identities of students by creating a mapping connection between the students' pseudonyms and real identities [39].

*5.2.4 Cost*
Another challenge concerns the cost involved in blockchain transactions, as managing and storing large amounts of student data on the blockchain can increase mining costs. Hence, it is necessary to manage development and operational costs of using blockchains in traditional education systems [24].

*5.2.5 Scalability*
Scalability concerns how the increasing number of participants and assets as well as larger transaction sizes can impact the blockchain applications' access latency. When addressing large sets of data and metadata, particularly regarding education, scalability should be dealt with in the preliminary architecture. It is difficult to foresee the direction and extent of blockchain technology considering potential future adaptations and new implementations for scalability [12].

*5.2.6 Blockchain platforms*
It is thus necessary to determine whether it is better to store the data on the blockchain or encrypt it and store the decryption keys on the blockchain. Existing blockchain platforms, including Bitcoin and Ethereum, are able to process approximately 10–15 transactions per second on average. Centralised applications used by credit card companies such as Visa can process approximately 5,000 to 8,000 transactions per second on average.

Further, as miners are more inclined towards high gas transactions (the highest-price-first-served model), transactions costs can increase.

*5.2.7 Consensus algorithms*
Because sensitive education information is stored in the blockchain, developing a secure blockchain system is a vital research area. Several researchers have focused on developing consensus algorithms for safeguarding transaction data, including Proof of Stack (PoS) and Proof of Work (PoW). Using blockchain technology in education may also include the development of consensus algorithms with the most benefit [24] [39].

*5.2.8 Motivation*
Another challenge is a lack of motivation. It is difficult to encourage stakeholders to implement blockchain-based applications in the traditional systems of education that have now been in place for a considerable time. Thus, further studies on blockchain usability for individuals are needed. Blockchain adoption can be improved in the education field through the development of usable blockchain-applications [13][22][24].

# 6. CONCLUSION
Blockchain is a distributed ledger technology that utilises cryptography techniques and distributed consensus algorithms for decentralisation, immutability, and traceability. The properties offered by blockchain and smart contracts can lead to several innovative applications in the context of higher education. It is possible to use blockchain technology for education in various innovative ways besides managing diplomas and evaluating achievements. Blockchain technology shows significant potential for learners and teachers in terms of its broad application for the design and implementation of learning activities, conducting formative evaluation and tracking the entire learning processes. However, there have been limited studies on this subject. Thus, evaluating the opportunities offered by how blockchain and smart contracts can be used to improve learning motivation and achievements is difficult. This systematic mapping review study aims to address this issue, by illuminating existing blockchain applications for higher education and highlighting the research challenges involved in implementing blockchain technology. The study results can help future researchers to identify and resolve additional challenges.